\documentclass[letterpaper, 10 pt, conference]{ieeeconf}  

\IEEEoverridecommandlockouts                              

\usepackage{graphicx}
\usepackage{url}

\usepackage{tabularx,booktabs}

\title{\LARGE \bf
A tidal lung simulation to quantify lung heterogeneity \\
with the Inspired Sinewave Test
}

\author{Minh~C.~Tran$^{1,2}$, 
			Douglas~C.~Crockett$^{1}$, 
             Phi~A.~Phan$^{1}$, 
             Stephen~J.~Payne$^{2}$, 
             Andrew~D.~Farmery$^{1}$
\thanks{$^{1}$M. Tran, D. Crockett, P. Phan and A. Farmery are with Nuffield Division of Anaesthetics, University of Oxford, UK.}%
\thanks{$^{2}$M. Tran and S. Payne are with the Institute of Biomedical Engineering, University of Oxford, UK.}
}

\begin{document}

\maketitle
\thispagestyle{empty}
\pagestyle{empty}

\begin{abstract}

We have created a lung simulation to quantify lung heterogeneity from the results of the inspired sinewave test (IST). The IST is a lung function test that is non-invasive, non-ionising and does not require patients’ cooperation. A tidal lung simulation is developed to assess this test and also a method is proposed to calculate  lung heterogeneity from IST results. A sensitivity analysis based on the Morris method and linear regression were applied to verify and to validate the simulation. Additionally, simulated emphysema and pulmonary embolism conditions were created using the simulation to assess the ability of the IST to identify these conditions. Experimental data from five pigs (pre-injured vs injured) were used for validation. This paper contributes to the development of the IST. Firstly, our sensitivity analysis reveals that the IST is highly accurate with an underestimation of about  $5\%$ of the simulated values. Sensitivity analysis suggested that both instability in  tidal volume and extreme expiratory flow coefficients during the test cause random errors in the IST results. Secondly, the ratios of IST results obtained at two tracer gas oscillation frequencies can identify lung heterogeneity ($ELV_{60}/ELV_{180}$ and $Qp_{60}/Qp_{180}$). There was dissimilarity between simulated emphysema and pulmonary embolism ($p<0.0001$). In the animal model, the control group had $ELV_{60}/ELV_{180}= 0.58$ compared with $0.39$ in injured animals ($p<0.0001$).
\newline

\indent \textit{Keywods}--- inspired sinewave test, lung simulation, lung heterogeneity.
\end{abstract}

\section{INTRODUCTION}

Respiratory diseases, including chronic obstructive pulmonary disease (COPD) and acute respiratory distress syndrome (ARDS) account for considerable mortality worldwide \cite{who}. Lung heterogeneity is proposed as a sensitive index of lung disease \cite{Cressoni2014, Bruce2018c, Mountain2017}. The lung heterogeneity index is a number describing the mismatch between ventilation, alveolar volume and pulmonary blood flow throughout the lung.  \cite{West2012}. This index has been mainly measured by invasive methods or medical imaging with ionising tracer gases \cite{Wagner2012}.

The inspired sinewave test (IST) is a non-invasive, non-ionising method, using an inert tracer gas  ($N_2O$) which is forced to oscillate at a given frequency to measure various cardio-pulmonary indices  \cite{Bruce2019, Crockett2019, Phan2017a}. The IST has been described  elsewhere  \cite{Phan2015}. The IST can measure deadspace volume ($V_D$), alveolar lung volume ($V_A$) in the form of effective lung volume ($ELV$) and pulmonary blood flow ($Qp$). The IST has the potential to detect lung disease at the early states based on lung heterogeneity indices.

We have therefore developed a lung simulation to investigate the ability of IST to determine lung heterogeneity. We propose a method to quantify lung heterogeneity by using two different forcing periods of the tracer gas for the IST. This method is validated by using both simulated data(three virtual patients - healthy, emphysema and pulmonary embolism) and ARDS animal data (pre - injured and post - injured).

\section{METHOD}
\subsection{Computational model}

\begin{figure}[tb]
	\centering
	\includegraphics[width =1\linewidth]{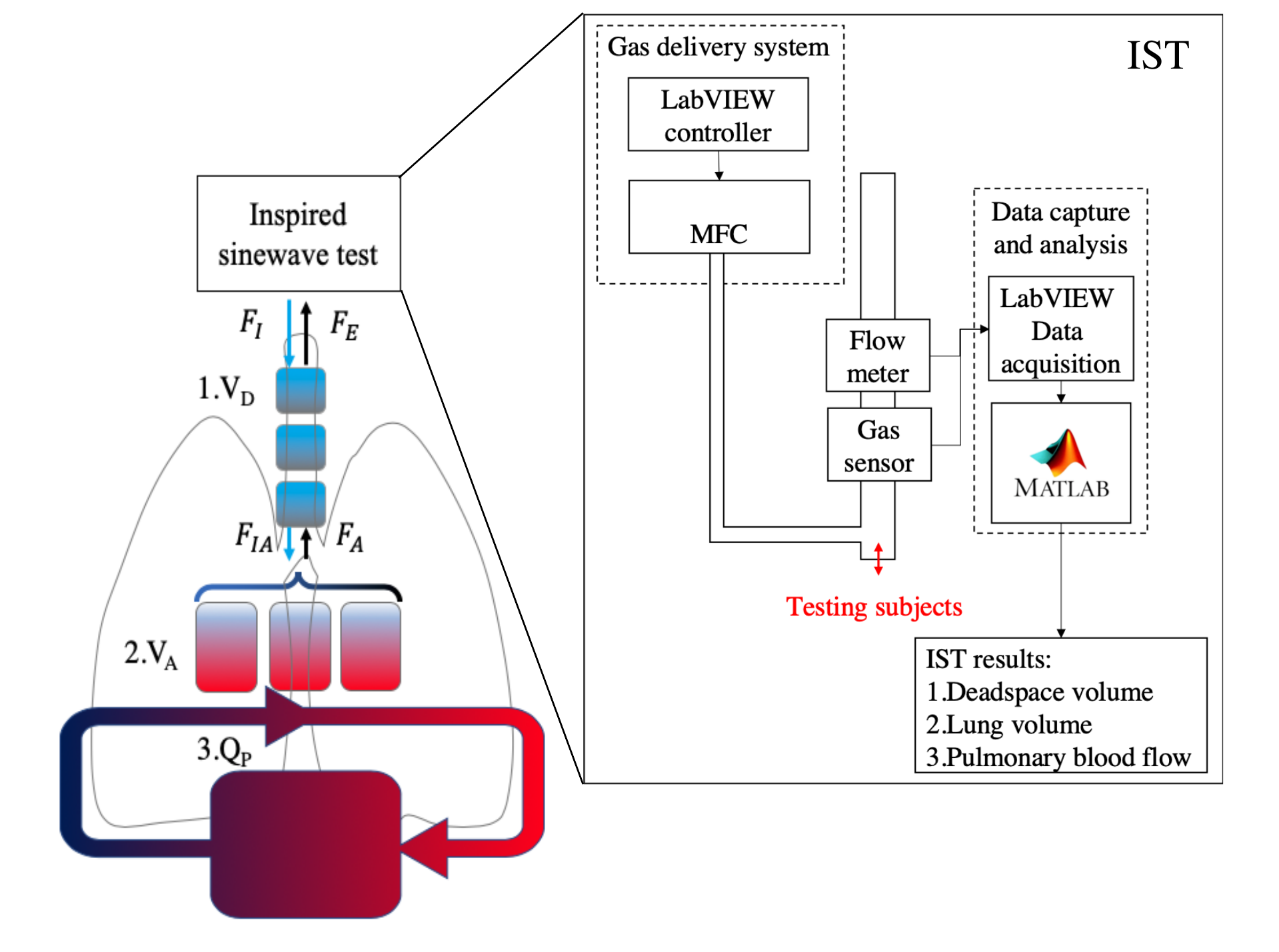}
	\caption{Schematic diagram of the tidal lung simulation and representation of the inspired sinewave test.}
	\label{fig:diagram}
\end{figure}

The lung simulation consisted of ten deadspace compartments \cite{harrison2017modelling}, three tidal lung compartments, a shunt and five body compartments \cite{Hahn2003}. Compartments were governed by conservation of mass equations. The \textit{ode45} solver and \textit{Simulink} from Matlab were adopted for simulation. The simulation was then combined directly with the IST analysis to create a forward-backward model \cite{Phan2015}. The model details are included in Appendix A and the schematic diagram of the model is shown in Fig \ref{fig:diagram}. 

For sensitivity analysis, the three lung compartments were assumed to be homogeneous and to behave as one well-mixed compartment. The setting parameters and testing ranges for the simulation are presented in Table 1.

The Morris method for sensitivity analysis was applied to verify and to validate the model \cite{Morris1991}. This method calculates the elementary effect values of each input parameter which were then used to identify the main input parameters and their relationship between the input and output \cite{Morris1991, Menberg2016}. A thousand samples were simulated for each parameter. Gaussian white noise was added to the simulated signal to justify the robustness of the IST results to noise.

\subsection{Method to quantify IST lung heterogeneity indices}
To assess the ability of the IST to calculate lung heterogeneity, heterogeneous lungs were firstly simulated according to the parameters in Yem et al. \cite{Yem2006}. These parameters were calculated from patients. Three conditions with different degrees of heterogeneity were simulated: normal lung, emphysema (ventilatory heterogeneous distributed), and pulmonary embolism (perfusion heterogeneous distributed). The IST was repeated twice with tracer gas oscillating at time periods of $180s$ and $60s$. The IST Ventilation heterogeneity of the lung was calculated by:

\begin{equation}
	ELV_{60}/ELV_{180}
\end{equation}
and the IST perfusion heterogeneity was calculated  by:
\begin{equation}Qp_{60}/Qp_{180}\end{equation}

Afterward, experimental data collected from five porcine models were analysed. The animals were under mechanical ventilation (control), before the lung was damaged by saline lavage to simulate the ARDS (injured) \cite{Lachmann1980}. The $180s$ and $60s$ IST were measured and compared during both states: pre-injured (control) and injured. The animal details, protocol and regulation are included elsewhere \cite{Crockett2019}.

\section{RESULTS}

According to the Morris method, the input parameters of the simulation were linearly linked to the IST outputs, as shown in Table 2. So, the lung simulation model is compatible with the IST. Unsurprisingly, alveolar volume input was the first rank in the most influential parameter to the $V_A$ output. Tidal volume and expiratory flow coefficients had high ranks, but they caused random error to the IST outputs because of their non-monotonic relationship to the output. In Table 2, when the input changed, recovered values were compared with the true values and recorded in the error (\%) column. Overall, IST errors were less than $5\%$. 

\begin{figure}[tb]
	\centering
	\includegraphics[width =1.1\linewidth]{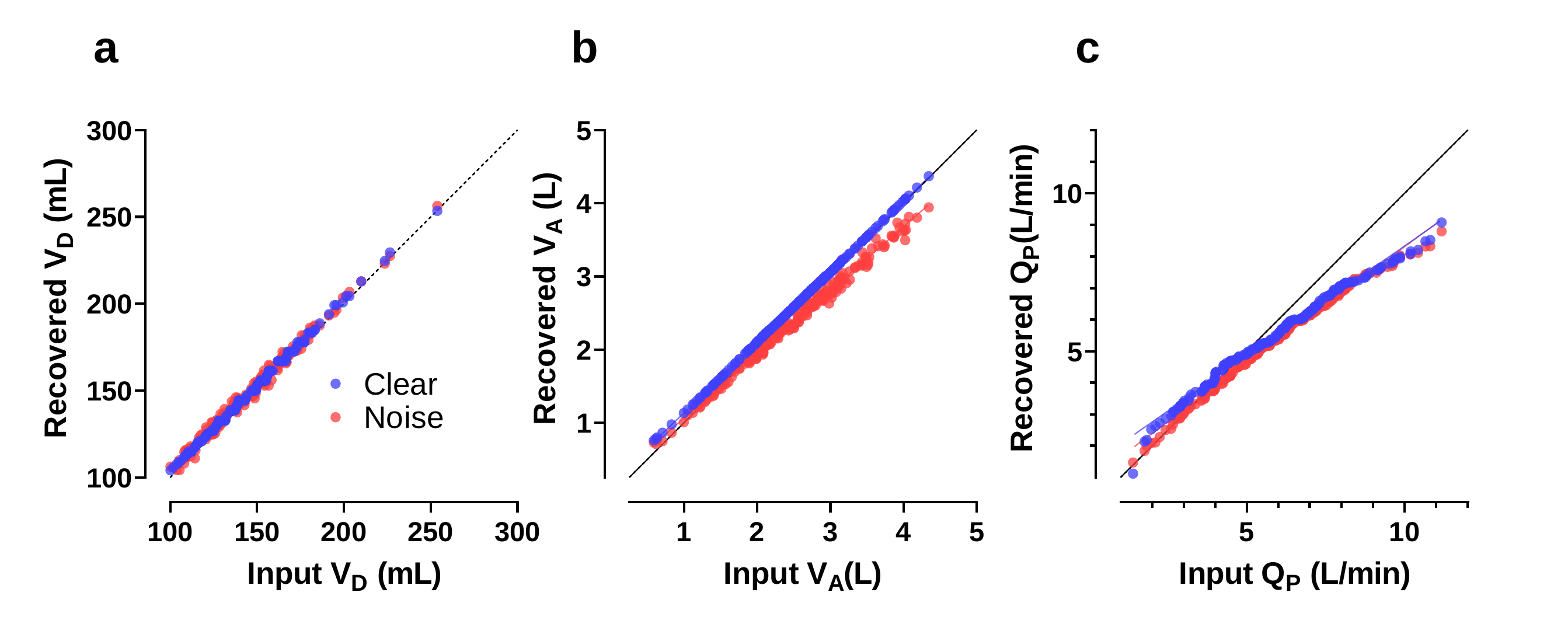}
	\caption{Linear regression analysis of the IST input and recovered values of $V_D$, $V_A$ and $Qp$. Gaussian white noise was applied.}
	\label{fig:A1}
\end{figure}

Fig. \ref{fig:A1} shows the results of linear regression of three output parameters. A strong correlation between the simulation inputs and the IST outputs ($R^2=1$) could be seen from the figures. The IST results also suggested that $V_D$ could be calculated correctly even with added noise. However, under noisy conditions, the $V_A$ regression slope shifted from $0.96$ to $0.87$ and the $Qp$ slope shifted from $0.70$ to $0.74$.

\begin{figure}[tb]
	\centering
	\includegraphics[width =0.85\linewidth]{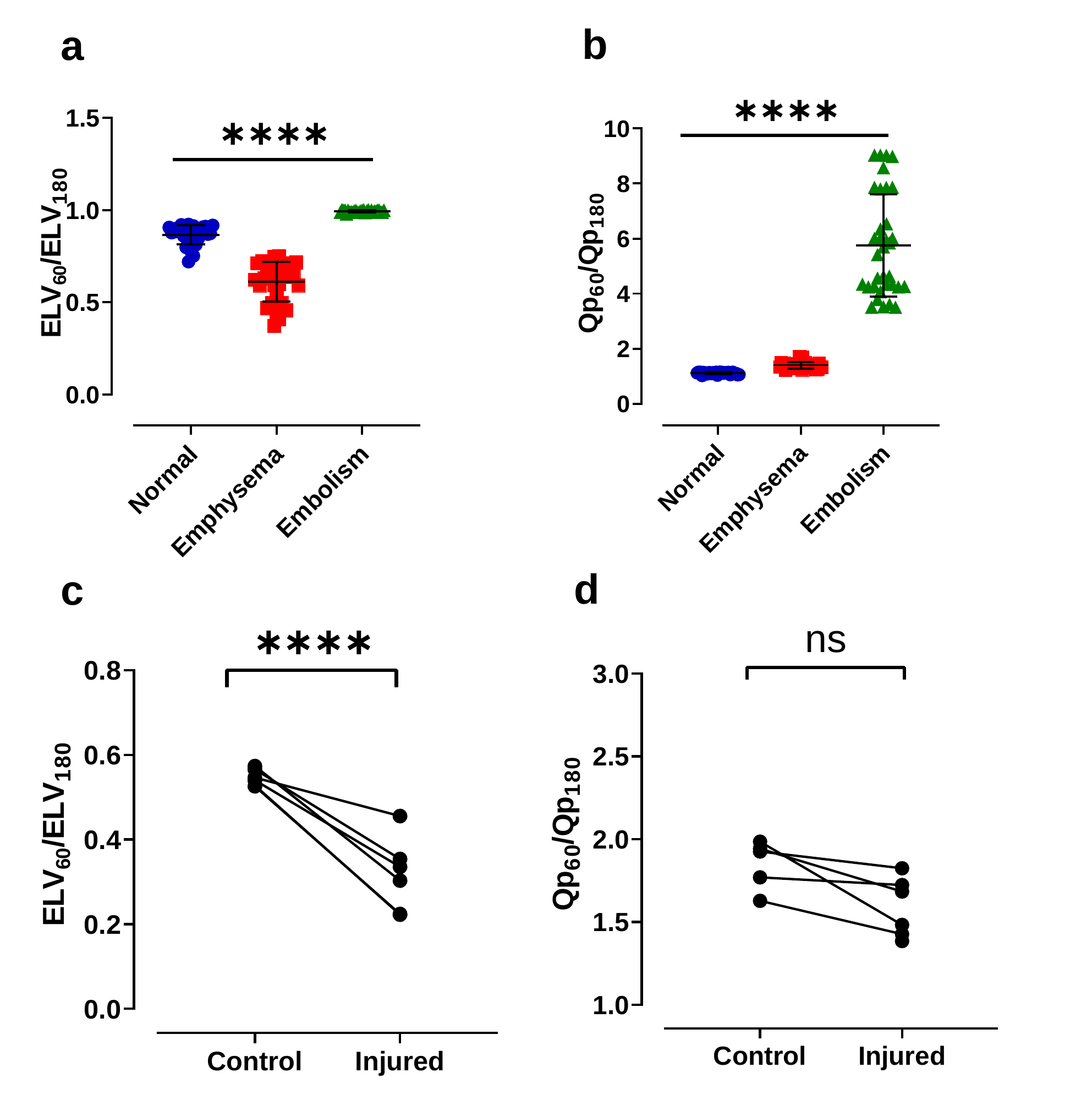}
	\caption{IST ratios of $ELV$ and $Qp$ with different lung conditions. In panel a and b are simulated data and analysed with the  ordinary one-way ANOVA test. Panel c and d are IST ratios between pre-injured (control group) and post-injured in five anesthetised pigs ($5 cmH_2O$ PEEP, tidal volume of $10mL/Kg$). Student’s t-test was implemented. ****$ = p<0.0001$, ns = non-significant. }
	\label{fig:A2}
\end{figure}

Diseased and healthy simulated patients could be distinguished easily using the different IST ratios, Fig. \ref{fig:A2}, panel a and b ($p<0.0001$). In emphysema, the ratio of $ELV_{60}/ELV_{180}$ could identify the heterogeneity of the lung (about $0.58$). However, the opposite was observed in the embolism. The ratio of $Qp_{60}/Qp_{180}$ was approximately $5.8$, higher than the values in the healthy and emphysema conditions.

Additionally, from the experiments in five animals (weight of $28 \pm 2kg$), the IST could classify the pre-injured (control) and the post-injured animals (ARDS model) reliably using the IST ventilation heterogeneity ratio ($p<0.0001$)(Fig. \ref{fig:A2}c). However, the classification with the IST perfusion heterogeneity ratio was not significant (cf. Fig. \ref{fig:A2}d).



\begin{table}[htb]
	\caption{Lung simulation parameters in normal lung, emphysema and embolism conditions \cite{Yem2006}. Mean + SD parameters in healthy conditions are ranges for the sensitivity analysis test.} 
	\vspace{-0.0cm}
	\centering
	\scalebox{0.87}{
		
		\begin{tabular}{l c c r } 

			\hline 
			\textbf{Variable}  &\textbf{Normal}    &\textbf{Emphysema}   &\textbf{Embolism}  \\
			\hline 
		
			\hline 
Deadspace volume (L)        &$0.15 \pm 0.025$                &$0.33$            &$0.157$  \\
Alveolar volume (L)          &$2.5 \pm 0.75$                  &$4.5$            &$2.813$  \\
Tidal volume (L)            &$0.7 \pm 0.3$                &$0.54$            &$0.46$  \\
Inhaled time (s)             &$2 \pm 0.5$                &$1$            &$0.75$  \\
Respiration rate \\(Breath/min)  &$12 \pm 5$           &$22$   &$29$ \\
Expiratory flow \\ coefficient (1/s)  &$1 \pm 0.3$                   &$1$            &$1$ \\
		Shunt (L/min)          &$0$                        &$0.015$            &$0.74$  \\
Blood flow (L/min)   &$5 \pm 1.5$                 &$5$            &$3.7$  \\
IST forced periods (s)      &$60$ \& $180$                    &$60$ \& $180$            &$60$ \& $180$  \\
Tidal fraction                &[$1$]                     &$[0.102$ $0.065$ $0.833]$            &$[0.334$ $0.666]$  \\
Lung volume fraction      &[$1$]                     &$[0.196$ $0.577$ $0.227]$            &$[0.333$ $0.667]$  \\
Blood flow fraction           &[$1$]                     &$[0.889$ $0.072$ $0.039]$            &$[0.507$ $0.493]$  \\

\hline 

		\end{tabular}
	}
	\label{table:cyc_ana} 
\end{table}
%

\begin{table*}[htb]
	\caption{Morris method results of the relationship between simulation parameters and IST results ($V_D$, $V_A$ and $Qp$).} 

	\centering

	\begin{tabular}{*{10}{l | c c c | c c c | c c c}}
		\hline		
		&  \multicolumn{3}{c}{\textbf{VD}}  &\multicolumn{3}{c}{\textbf{VA}}    &\multicolumn{3}{c}{\textbf{Qp}}    \\
		\hline

	& \textbf{Rank}  &\textbf{Relationship}   &\textbf{Error (\%)}   &\textbf{Rank}  &\textbf{Relationship}  &\textbf{Error (\%)}   &\textbf{Rank}  &\textbf{Relationship}   &\textbf{Error (\%)} \\			
\hline 
Deadspace volume (L)       &\textbf{2} &\textbf{L}  &\textbf{3} &7 &AM &-  &9 &M &-  \\
Alveolar volume (L)        &$3$ &M  &$-2$&\textbf{1} &\textbf{L}  &\textbf{-7} &4 &M &4  \\
Tidal volume (L)           &1   &NM &$-3$&2 &NM &-5 &1 &AM &-4  \\ 
Inhaled time (s)           &5   &NM &-   &9 &NM &-  &7 &NM &-  \\
RR (Breath/min)            &8   &NM &-   &6 &M  &-  &8 &AM &-  \\
Expiratory flow  (1/s)   &4   &NM &$-4$  &3 &NM &-2 &3 &NM &1  \\
Shunt (L/min)              &6   &NM &-   &4 &M  &-4 &5 &AM &4  \\
Bloodflow (L/min)          &5   &NM &$3$   &5 &M  &-3 &\textbf{2} &\textbf{L}  &\textbf{4}  \\
IST forced period (s)          &7   &AM &-   &8 &L  &-  &6 &AM &-  \\
\hline 
\multicolumn{10}{l}{L: linear, M: monotonic, AM: almost monotonic, NM: non monotonic. Percentage error of some input of some parameter were negligible.} \\
\multicolumn{10}{l}{Error (\%) was calculated by percentage difference between simulation input and the recovered values} \\

\hline  

	\end{tabular}

\end{table*}

\section{DISCUSSION}

Lung simulations can be used to validate lung measurement techniques such as the IST and provide information to improve performance. This work has combined a validated lung simulation using parameters taken from the literature to undertake error analysis and evaluation of the accuracy and precision of the IST. We empirically demonstrated that the IST ratio of two frequencies is an impactful method to quantify the lung heterogeneity using only IST results.

The results of sensitivity analysis and linear regression also suggested that the IST analysis offered a high accuracy (error of 3\% to 5\%) and a high precision (6\% variation) in a healthy lung condition. 

The results of this work are consistent with previous comparison studies and provide further understanding of the IST. Previous work comparing the IST measurement of $Qp$ and $Qp$ measured from the thermodilution showed good agreement, although the IST was seen to marginally underestimate true $Qp$   \cite{Bruce2019}. A similar result has been shown using our simulation, where IST $Qp$  was seen to underestimate the true value by $0.52L/min$. Furthermore, our simulations show that the error will increase by up to 25\% of the real value as $Qp$ exceeds $7L/min$,  suggesting that in its current embodiment, the IST is best suited to normal or low cardiac output states.

Secondly, previous clinical and animal studies comparing IST estimates of $V_A$ to those measured by plethysmography found that the IST underestimated  lung volume by an amount ranging from $0.88$ to  $1.09L$ compared to CT and plethysmography  \cite{Bruce2018c, Crockett2019}. In our simulation however, IST ELV underestimated $V_A$ by only $0.2L$. One explanation for this is that plethysmography and CT measures estimate the total volume of lung gas, including trapped gas, whereas the IST measure the volume involved in gas exchange.  Our simulation did not include any such trapped gas compartment.  Further underestimation of $V_A$ was discovered by our sensitivity analysis which showed that a changing tidal volume during IST (as wold occur during real breathing) caused errors which contributed to the underestimation of $V_A$.

This is the first study of the lung heterogeneity measured entirely by the IST and to evaluate its values in simulated and experimental data. Bruce et al. \cite{Bruce2018c} showed that ventilation heterogeneity can be captured by the IST and the plethysmography. Our work only implemented IST results at 60s and 180s signals and required no additional measurement technique. It could be a potential parameter to identify lung disease at early stages. Although, there are limitations of the IST ratios, further research could be done in the future.

Modelling saves time, money and effort. In this work, thousands of simulated patients with different lung conditions were created and tested by the IST. We provided a simple and effective lung simulation that combines the advantages of previous simulations and works directly with a clinical measurement (IST).

The IST is suitable for bedside monitoring. It is non-invasive, easy to perform and does not require patient effort. Furthermore, our analysis has shown that when keeping the tidal volume unchanged, the IST error was minimised. Thus, it may be ideally suited to patients undergoing mechanical ventilation.

\section{CONCLUSION}

In this paper, we have used a mathematical model to create a lung simulation which is compatible with the IST. The results suggested that stable tidal volume can reduce error when performing the IST. We also have demonstrated how the IST can be used to classify lung heterogeneity. The method was applied in animal data (control vs ARDS injured). We proposed two new indices of heterogeneity; one describing the mismatch between regional ventilation and regional volume, the other describing the mismatch between regional ventilation and perfusion.  These are alternative parameters to those measured by sophisticated invasive methods or medical imaging with ionising radiation and/or tracers.

\section{Appendix A}
The equation for the deadspace compartment during inspiration is:
\begin{equation}
V_{M,i} \frac{dF_{M,i}}{dt} = \dot{V}_A(t) (F_{M,i-1}(t) - F_{M,i} (t)) 
\end{equation}
where $i$ is the number of deadspace compartments (from 1 to 10); and $F_{(M,1)} (t)=F_I (t)$ and $F_{M,10} (t)=F_{IA} (t)$. $V_{M,i}$ is the deadspace volume of a compartment $i$ and $\dot{V}_A (t)$ is the flow at time $t$. The equivalent equation during expiration is:
\begin{equation}
V_{M,i} \frac{dF_{M,i}}{dt} = \dot{V}_A (t) (F_{M,i}(t) - F_{M,i+1} (t)) 
\end{equation}
where $i$ is from 1 to 10 and $F_{M,1} (t)=F_E (t)$ and $F_{M,10} (t)=F_{A} (t)$. $F_I$, $F_{IA}$  and $F_E$ are the fractional concentrations of the tracer gas coming into the deadspace compartment, coming out from the deadspace compartment to the lung compartment and exhaling from the deadspace to the environment, respectively. 

For the tidal compartment during inspiration and expiration:
\begin{equation}
V_A (t) = \bar{V}_A + \frac{V_T \times t}{t_i}
\end{equation}

\begin{equation}
V_A (t) = \bar{V}_A + V_T \exp({-\gamma (t - t_i)})
\end{equation}

where $\gamma$ is the rate-constant of expiratory flow \cite{harrison2017modelling}, $V_A (t)$ is the alveolar volume, $V_T$ is the tidal volume and $t_i$ is the inspiration starting time. $\bar {V}_A$ is end-expired alveolar volume. All compartments are linked together by governed equations: during inspiration is (\textit{7}) and expiration is (\textit{8}):
\begin{equation}
\frac{d}{dt}(V_{A}(t) \times F_{A}(t)) = \dot{V}_A (t)\times F_{IA}(t) + \dot{Q}_{p} (C_{\bar{v}} - C_{a})
\label{goevnInspi}
\end{equation}

\begin{equation}
\frac{d}{dt}(V_{A}(t) \times F_{A}(t)) = \dot{V}_A (t)\times F_{A}(t) + \dot{Q}_{p} (C_{\bar{v}} - C_{a})
\label{goevnExpi}
\end{equation}

To create lung heterogeneity, the fraction of ventilation ($ \dot{V}_A (t)$), lung volume ($\bar{V}_A$) and perfusion ($\dot{Q}_{p}$) changed according to Table 1.  $C_{\bar{v}}$  and $C_a$  are the mixed venous and pulmonary end-capillary gas concentrations. The body compartments of a standard human consist of five different tissues inside the human body with different tissue-gas coefficient \cite{Hahn1993, farmery2008interrogation}.

\begin{equation}
V_{i}^{\ast} \frac{dC_{i}(t)}{dt} = \dot{Q}_{i} (C_{\bar{a}}(t) - C_{i}(t))
\end{equation}
where  $ V_{i}^{\ast}$, $C_i$ and $\dot{Q}_{i}$ are the equivalent blood volume, concentration of the tracer gas and blood flow rate of $i$ compartment. $V_{i}$  is given by the following equation where $V_{(b,i)}$ and $V_{(t,i)}$ are the blood and tissue volumes of each body compartments and $\lambda_{(t,i)}$ is the Ostwald tissue-gas coefficient of each compartment. The values of $V_{(b,i)}$, $V_{(t,i)}$ and $\lambda_{(t,i)}$ are given in data of a $70kg$ standard man  \cite{farmery2008interrogation}.

\begin{equation}
V_{i}^{\ast} = V_{b,i} + \frac{\lambda_{t,i}}{\lambda} V_{t,i}
\end{equation}

Furthermore, the Ostwald coefficient is the blood-gas partition coefficient ($\lambda$). 
In the IST, the Ostwald coefficient of the $N_2O$ is $0.47$. The body temperature is assumed to be constant ($37 ^oC$), so the Ostwald coefficient is stable.

\begin{equation}
C_{a} = \lambda \times F_{A}
		\label{equ_lamda}
\end{equation}

Finally, the shunt equation is:
\begin{equation}
\frac{\dot{Q}_{S}}{\dot{Q}_{T}}  = \frac{C_{a}(t) - C_{\bar{a}}(t) }{C_{a}(t) - C_{\bar{v}}(t)}
\end{equation}


\bibliographystyle{IEEEbib}
\bibliography{EMBC_Minh_v08}


\end{document}